# Quantum Lithography Based on Multiple Light Exposures for Arbitrary Sub-Diffraction-Limit Pattern Generation

Hee Su Park and Sun Kyung Lee

Korea Research Institute of Standards and Science (KRISS),

Daejeon 305-340, Republic of Korea

We propose a super-resolution quantum lithography scheme based on coherent population trapping in $\Lambda$-type atoms coupled to temporally-cascaded standing-wave driving fields. By realizing effective multiplication of optical intensity profiles on an atomic state density distribution, the scheme enables an arbitrarily high degree of resolution enhancement without modifying the atomic level structure of the photographic medium or the light source configuration. It is also shown that the visibility of the super-resolution patterns is preserved under a significant atomic state decoherence rate.

42.50.Ct, 42.50.Gy, 42.50.St, 85.40.Hp

Improving the spatial resolution is a key issue in developing lithography technologies. Besides developing shorter-wavelength optical components to improve the diffraction-limited resolution, various schemes have been proposed to beat the diffraction limit without changing the wavelength of light [1-14], some of which have been experimentally demonstrated [2-4,7,10,12-14]. Most of the proposed schemes incorporate photographic media with multi-photon absorption ability to reduce the effective wavelength of light that can be either a classical electromagnetic field [1-6] or quantum mechanically correlated photons [9-14]. To date this necessity of multi-photon absorbers has caused limited visibility of the fringe patterns due to noise by absorption of lower-number of photons and insufficient source intensity. In contrast, some proposals have avoided this necessity by relying on the coherent population trapping (CPT) mechanism [15,16], which reveals a sub-diffraction-limited atomic density distribution [7,8,17]. Although such CPT-based techniques can realize super-resolution with lower-intensity lasers, experiments in solid states have not yet been reported seemingly due to the difficulty of maintaining a sufficient level of atomic coherence [16]. In this paper, we propose a novel CPT-based scheme with a simple $\Lambda$-type atomic system and three distinct classical light sources. Unlike the previously proposed or demonstrated schemes, our scheme achieves arbitrarily high resolution enhancement without modifying photographic media or sacrificing visibility; in addition, a high level of visibility is maintained even under a significant atomic-state decoherence. The key idea is sequentially applying phase-shifted optical standing waves for CPT and an optical quenching operation [17] $N$ times, by which the final population of a ground state becomes proportional to the multiplication of $N$ standing wave distributions. The degree of resolution enhancement $N$ can arbitrarily increase according to the number of exposure steps, although reduction of the maximum state density has to be considered for developing processes after the light exposure.

The method to generate fringe patterns that show $N$-times greater spatial frequency than ordinary two-beam interference fringes is schematically shown in Figs. 1(a)-(c). Two signal fields and one ancilla field are incident on a substrate that possesses three stable ground states ($g_1$, $g_2$, $q$) and two excited states ($e_1$, $e_2$). The atomic levels are in resonance with the optical fields as shown in Fig. 1(b), the excited state $|e_1\rangle$ spontaneously decays to the ground states $|g_1\rangle$ and $|g_2\rangle$, and the excited state $|e_2\rangle$ decays to the ground states $|g_2\rangle$ and $|q\rangle$. Here the ground state $|q\rangle$ is not necessarily a pure state, but can be a mixture of ground states that are not coupled to the other states by the applied fields. The unit exposure process comprises steps A and B. In step A, the signal fields that contain spatial intensity distribution are exposed to the substrate. In step B, the ancilla field is exposed with a uniform spatial distribution. The substrate atoms are initially at $|g_1\rangle$, and the state will evolve to the dark state comprising $|g_1\rangle$ and $|g_2\rangle$ in step A as

$$|D_A\rangle = \left(S|g_1\rangle - R|g_2\rangle\right) \Big/ \sqrt{|S|^2 + |R|^2} \;, \quad (1)$$

where $S$ and $R$ correspond to the Rabi frequencies $\mu \cdot E/2\hbar$ of the signal 1 and signal 2 fields, respectively, with $\mu$ being the electric dipole moment and $E$ the electric field amplitude. In step B, the ancilla field transfers all the atoms from $|g_2\rangle$ to $|q\rangle$, and the final state becomes a mixture of $|g_1\rangle$ and $|q\rangle$ with the population at $|g_1\rangle$ being $|S|^2/(|S|^2+|R|^2)$. If $S$ and $R$ are two equal-amplitude standing waves with a relative phase shift of $\pi/2$, $|S|^2/(|S|^2+|R|^2)$ becomes $[1-\cos(2k_0z+\phi)]/2$ [8], which reveals a sinusoidal spatial modulation of the state density. Here $k_0$ is the $z$-component wave number and $\phi$ is a constant phase tunable with the phase shifters in Fig. 1(a). If the unit exposure process is repeated $N$ times with the phase $\phi = 2\pi(\nu-1)/N$ for the $\nu$-th process, the state density profiles in the unit processes are multiplied to yield a spatial density modulation of the final state $|\Psi_{final}\rangle$

$$\left|\langle g_1 | \Psi_{final}\rangle\right|^2 = \prod_{\nu=1}^{N} [(1-\cos(2k_0z + 2\pi(\nu-1)/N))]/2 = \frac{1}{2^N}\left(\frac{1-\cos(2Nk_0z)}{2}\right), \quad (2)$$

which reveals the fringes with an *N* times ordinary wave number. Such modulation can be further converted to a lithographic pattern through development sensitive to the $|g_1\rangle$ density; for example, by utilizing a method used in atomic resonance lithography in which non-uniformly distributed metastable atoms form a resist layer for masking chemical etching when they come into contact with vapor molecules [18,19] or self-assembled monolayers [20].

The proposed method can be generalized to realize a more complex fringe pattern by adjusting the phase and the fringe visibility in each step. A standing wave *S* with a fringe visibility of < 1 can be generated in step A by counterpropagating signal 1 beams with unequal intensities. The corresponding standing wave *R* by signal 2 beams must have a 100% visibility and the amplitude that matches with the ac component of *S*. With the generalized phase and visibility, the final state density is rewritten as

$$\left|\langle g_1 | \Psi_{final} \rangle\right|^2 = \prod_{\nu=1}^{N}\left[\frac{1 + r_\nu e^{2ik_0 z} + r_\nu^* e^{-2ik_0 z}}{1 + 2|r_\nu|}\right] = \prod_{\nu=1}^{N}\frac{1}{(1 + 2|r_\nu|)} \cdot \sum_{\mu=0}^{N}\left(f_\mu e^{2i\mu k_0 z} + c.c.\right), \quad (3)$$

where $r_\nu$'s are complex constants with $0 \leq |r_\nu| \leq 1/2$ and *c.c.* stands for complex conjugate. Here the complex Fourier coefficients $f_\mu$'s are derived from $r_\nu$'s as

$$2f_0 = 1 + \sum r_{\nu_1} r_{\nu_2}^* + \sum r_{\nu_1} r_{\nu_2} r_{\nu_3}^* r_{\nu_4}^* + \cdots,$$
$$f_1 = \sum r_{\nu_1} + \sum r_{\nu_1} r_{\nu_2} r_{\nu_3}^* + \sum r_{\nu_1} r_{\nu_2} r_{\nu_3} r_{\nu_4}^* r_{\nu_5}^* + \cdots,$$
$$\vdots \quad (4)$$
$$f_N = r_1 r_2 \cdots r_N ,$$

where the indices $\nu_i$'s in the summation are all distinct. Equation (3) shows that the final state density profile has a form of an *N*-th order truncated Fourier series. Moreover, the number of variables $r_\nu$'s is less by one than the number of coefficients of a real-valued *N*- th order Fourier series, and therefore is sufficient to generate an arbitrary pattern given by the Fourier series within an ambiguity of overall amplitude. However, Eq. (3) obviously cannot generate a negative state density, so we numerically fitted the result of Eq. (3) with the target pattern and then compared it with the truncated Fourier series as follows rather than solving Eq. (4) with the Fourier coefficients for a target pattern.

The results of the numerical fitting with $N = 10$ for a square-shaped target pattern described in [6] and [9] are shown in Fig. 2. The target pattern and the trial pattern in Eq. (3) were compared at 20 points at $z = \pi/(20k_0) \cdot \nu$ ($\nu = -10, -9, \ldots, 9$) and the magnitude of difference between the normalized vectors $\vec{v}_{ta}/|\vec{v}_{ta}|$ and $\vec{v}_{tr}/|\vec{v}_{tr}|$ was minimized using a nonlinear least-squares algorithm, where $\vec{v}_{ta}$ and $\vec{v}_{tr}$ are the 20-element vectors comprising the target and trial patterns at the sampling points, respectively. According to the Nyquist-Shannon theorem, these two vectors become identical when the Fourier components given by Eqs. (3) and (4) match the 20-element truncated Fourier series calculated from the target pattern. Normalization of the two vectors is introduced in order to compare the patterns without considering the difference of the overall amplitudes. In Fig. 2, the fitted curve using Eq. (3) successfully demonstrates the same amount of proximity with the target pattern as the truncated Fourier series except in regions where the Fourier-analysis-based curve goes negative. The maximum state density $|\langle g_1|\Psi_{final}\rangle|^2$ at $z = 0$ was $3.86 \times 10^{-5}$ before normalization.

When the overall reduction of the state density of the above examples seriously deteriorates the lithography performance, a sub-diffraction-limited pattern with a high

state density can be achieved using a point-by-point patterning [21] with all $r_v$'s set as 1/2:

$$\left|\langle g_1 | \Psi_{final} \rangle\right|^2 = \prod_{v=1}^{N}\left[\frac{1+\cos(2k_0 z)}{2}\right] = \cos^{2N}(k_0 z) . \quad (5)$$

In this case, a single peak with the maximum state density of unity is created at $z = 0$ with no sidebands. This strategy using multiple exposures of identical fields can also be combined with the quenching method [17,22,23] for generating sub-diffraction-limited patterns. For example, in step A, one applies a spatially uniform signal 1 field together with a signal 2 field generating a standing wave whose average intensity is much greater than the signal field intensity. Then the state density of $|g_1\rangle$ is tightly localized near the node of the signal 2 field. Repeating the above exposure step together with step B further reduces the linewidth of the state density.

The experimental feasibility of this scheme was investigated based on the influence of atomic state decoherence on the final state density distribution [16]. Under nonzero decoherence rate, a part of atoms occupy $|e_1\rangle$ at the end of step A, and then decay to $|g_1\rangle$ in step B with a probability determined by the branching ratio of the spontaneous decay rates of $|e_1\rangle$ to $|g_1\rangle$ and $|g_2\rangle$. The state density distributions after a unit exposure step in Eq. (2) are plotted in Fig. 3(a), where all the three phase differences between $|g_1\rangle$, $|g_2\rangle$, and $|e_1\rangle$ decohere at a same rate $\gamma_d$. The branching ratio ($|e_1\rangle \to |g_1\rangle$):($|e_1\rangle \to |g_2\rangle$) was 1:1 and the total field intensity $|S|^2+|R|^2$ was $\Gamma^2$ in the calculations, where $\Gamma$ is the spontaneous decay rate of the excited state $|e_1\rangle$ to the ground states. Decoherence does not affect the visibility and the FWHM width under these conditions as shown in Fig. 3(a). In addition, applying smaller-amplitude fields further decreases the decoherence-induced distortion;

for example, the distortion becomes less than 1% when $|S|^2+|R|^2$ is $(0.1\Gamma)^2$. In contrast, under a finite decoherence rate, a branching ratio $\Gamma_2:\Gamma_1$ different from 1:1 changes the FWHM width of the density profile as shown in Fig. 3(b), where $\Gamma_1$ and $\Gamma_2$ are the decay rates of $|e_1\rangle$ to $|g_1\rangle$ and $|g_2\rangle$, respectively. Such change of the peak width can be corrected by applying $S$ and $R$ fields with unequal ac components although the overall pattern still differs from an ideal sine curve.

Methods to extend the scheme to an arbitrary $N$-th order 2D Fourier series are currently not straightforward; however, some super-resolution 2D patterns can still be generated with the setup in Fig. 1(a). For example, one can sequentially rotate the substrate to change the wave vectors of signal 1 and 2 beams as $\pm(k_0\cos\theta\cdot\mathbf{x} + k_0\sin\theta\cdot\mathbf{y})$, where $\mathbf{x}$ and $\mathbf{y}$ are the orthogonal unit vectors on the substrate plane. By applying multiple exposure steps for each value of $\theta$, super-resolution in any direction can be realized as shown in Fig. 4, where a C-shaped 2D target pattern is reproduced using 36 exposure steps. Angle $\theta$ was varied six times as $0, \pi/6, 2\pi/6, ..., 5\pi/6$ and each angle configuration comprised six unit exposure steps in the calculation. Least-square fit was performed with 50×50 samples and the final state density was $2.17\times10^{-8}$ at the maximum.

The high level of visibility shown in Figs. 2-4 originates from the fact that the state density at $|g_1\rangle$, where signal 1 vanishes, always becomes zero regardless of an atomic decoherence and that the zero state density is maintained throughout the following exposure steps. Therefore, the keys to achieving high visibility in practical implementation are the quality of the optical standing waves and resist materials that can work properly even when the maximum state density of the relevant atoms is less than unity. Conversion of the atomic state density profile to a mask pattern has been experimentally demonstrated with a selective carbon-containing resist formation using ground- and metastable-state Ar atoms [18,19]. However, it still remains a technological challenge to find a mechanism that is applicable to the solid-state CPT media such as neutral-donor($D^0$)-doped semiconductor (GaAs) [24] or Pr:SiO crystal [25]. The signal wavelength of 817 nm used in the recent experiment [24] will produce a fringe period of 41 nm with $N$=10 according to Eq. (2), which provides a reasonable advantage for lithography using visible or near IR light.

In summary, we have shown that the proposed lithography scheme based on CPT and optical quenching can enhance spatial resolution by an arbitrarily large number of times. The patterns encoded in the state density whose maximum is less than one can be effectively developed with the current techniques [18-19] if the number of atoms per unit volume is sufficiently large. Combination with the quenching-based super-resolution lithography techniques is also possible for further improvement of the spatial resolution. It is important to note that a decoherence rate comparable with atomic decay rates is permissible and that weak optical fields are more preferred to strong fields to generate ideal patterns. We believe these features will lessen the constraints of experimental realization in solid states compared to conventional CPT-based technologies.

This work was supported by the KRISS projects 'Single-Quantum-Based Metrology in Nanoscale' and 'Establishment of Research Foundation for SI Re-definition.' H. S. P. thanks D.-H. Lee for helpful discussions.

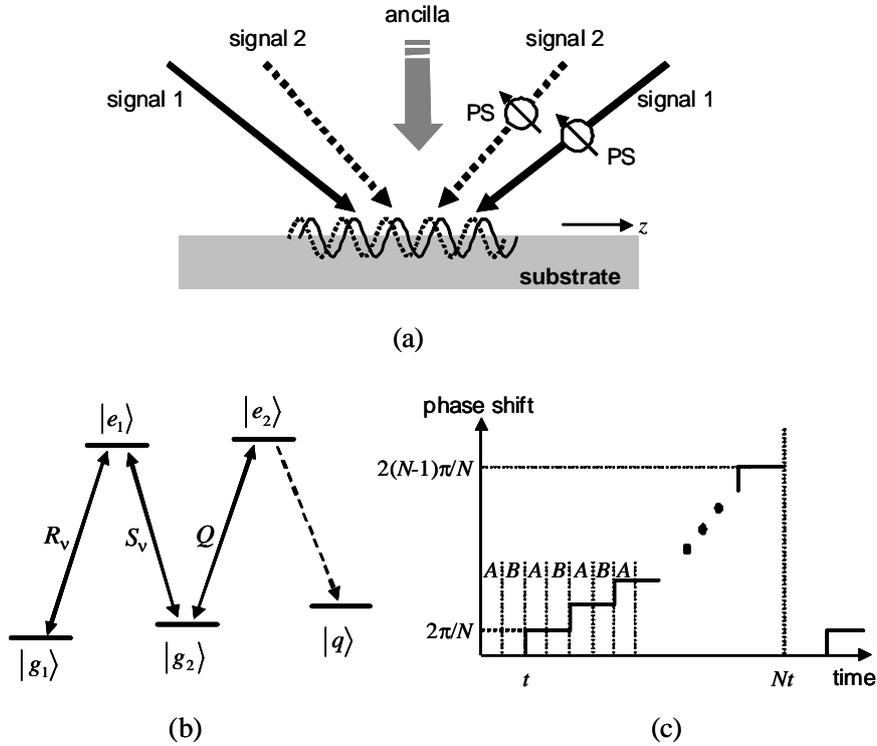

**Fig. 1. (a) Schematic of super-resolution lithography setup. Each of the signal 1 and signal 2 fields generates a standing wave that has an identical fringe amplitude and phase shifted by π/2 with each other. (b) Atomic level structure of the substrate. $S_v$, $R_v$, and $Q$ are the Rabi frequencies of the corresponding transitions that are driven by the signal 1, signal 2, and ancilla fields, respectively. (c) Sequence of the phase shift applied to the signal 1 and 2 fields for generating interferometric fringes with a period of $\pi/k_0 \cdot 1/N$. In step A, only the signal fields are exposed to the substrate, and in step B, only the ancilla field is turned on.**

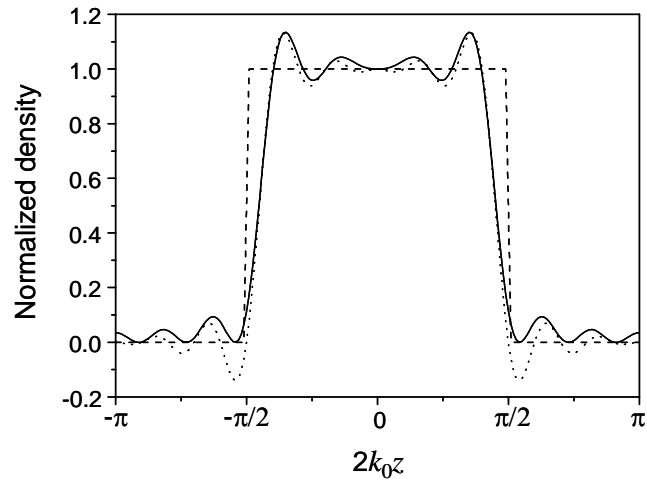

Fig. 2. State density profile (solid line) approximating the target pattern (dashed line) generated by the proposed lithography process with $N = 10$. For comparison, the 20-element truncated Fourier series with frequency components of $-10\cdot(k_0/\pi)$, $-9\cdot(k_0/\pi)$, ..., $9\cdot(k_0/\pi)$ is drawn with a dotted line.

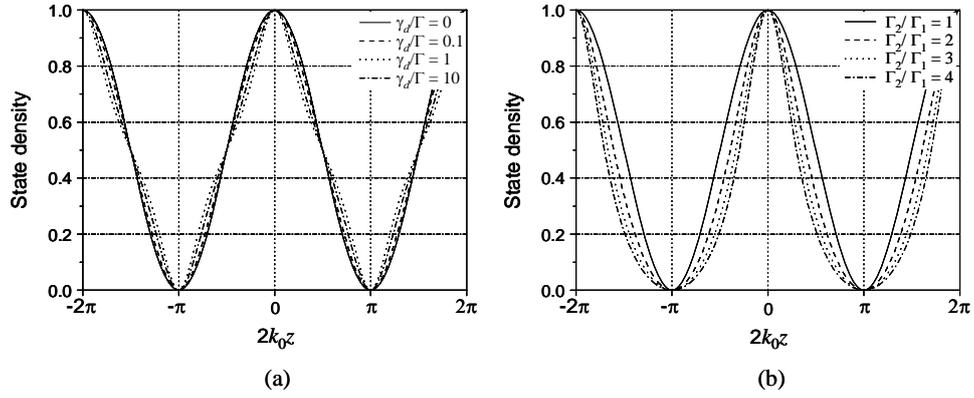

**Fig. 3. State density of $|g_1\rangle$ after a unit exposure step; (a) with varied decoherence rate $\gamma_d$ ($|S|^2+|R|^2 = \Gamma^2$, $\Gamma_1 = \Gamma_2 = \Gamma/2$), (b) with varied decay rate ratio $\Gamma_2/\Gamma_1$ ($|S|^2+|R|^2 = (0.1\Gamma)^2$, $\Gamma_1+\Gamma_2 = \Gamma$, $\gamma_d = \Gamma$).**

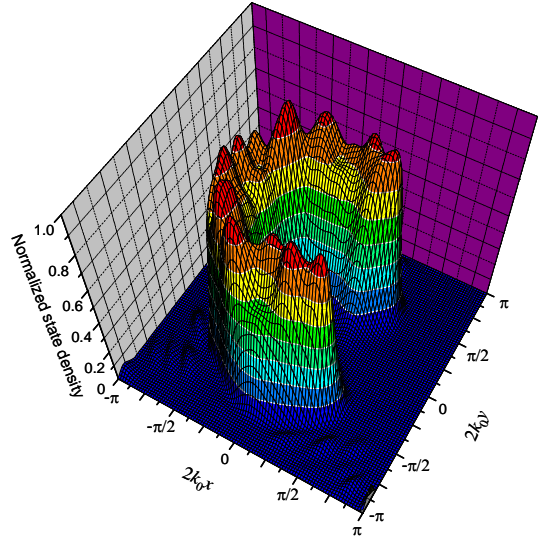

**Fig. 4.** Two-dimensional state density of $|g_1\rangle$ after 36 unit exposure steps. Standing wave vectors are aligned along $0, \pi/6, \ldots, 5\pi/6$ with respect to the *x*-axis and six exposure steps are performed with each angle configuration. The target pattern $f(r,\theta)$ is defined as $f(r,\theta) = 1$ for $\pi/3 < r < 2\pi/3$ and $\pi/4 < \theta < 7\pi/4$ and $f(x,y) = 0$ elsewhere, where $(r,\theta)$ is the polar coordinate on the substrate plane.